\documentstyle[12pt]{article}

\input epsf
\begin{document}

\def\req#1{(\ref{#1})}
\newcommand{\del} {\partial}
\newcommand{\ov} {\over}
\newcommand{\ha} {{1\over 2}} \newcommand{\fourth} {{1\over 4}}
\newcommand{\p} {\phi} \newcommand{\vp} {{\varphi}} \newcommand{\tvp} {{\tilde
\vp}}
\newcommand{\be}{\begin{equation}} \newcommand{\ee}{\end{equation}}
\newcommand{\ba}{\begin{eqnarray}} \newcommand{\ea}{\end{eqnarray}}
\newcommand{\la}{\label}
\newcommand{\vol}{\hbox{Vol}}
\newcommand{\lb}{\large\bf}
\newcommand{\bb} {\bibitem}
\newcommand{\np} {{\it Nucl. Phys. }} \newcommand{\pl} {{\it Phys. Lett. }}
\newcommand{\pr} {{\it Phys. Rev. }} \newcommand{\mpl} {{\it Mod. Phys. Lett.
}}
\newcommand\stwo{{\sqrt2}}
\newcommand\g{{{\hat g}e^\phi}}
\newcommand\bx{{\hbox{ $\sqcup$}\llap{\hbox{$\sqcap$}}}}
\newcommand\hatbx{{\hbox{ $\sqcup$}\llap{\hbox{$\hat\sqcap$}}}}
\newcommand\sq{{\sqrt g}}
\newcommand\sqhat{{\sqrt{\hat g}}}
\newcommand\sqphi{{\sqrt{\hat g}} e^\phi}
\newcommand\sqalpha{{\sqrt{\hat g}}e^{\alpha\phi}}

\hfill BUTP-94/14

\hfill hep-th/9412051

\begin{center}
 {\large\bf RG flow in $2d$ field theory coupled to gravity}
 \end{center}
\begin{center}
C. Schmidhuber\footnote{Talk at the 28th International Symposium on Elementary
Particle Theory
in Wendisch--Rietz, Germany, Aug 29 -- Sep 3, 1994.
Work supported by Schweizerischer Nationalfonds.}  \vskip1mm
 {\it Institute for Theoretical Physics, University of Bern, CH-3012 Bern,
Switzerland}
\end{center}
\vspace{.1truein}
\noindent {\bf Abstract:}
The renormalization group flow in two--dimensional field theories is modified
if they are
coupled to gravity. Beta function coefficients are changed, the $c$--theorem is
no longer strictly valid,
and flows from fixed points with central charge $c>25$ to fixed points with
$c<25$
are forbidden. This is discussed in general and at two examples, the
Kosterlitz--Thouless
phase transition and the Wess--Zumino--Witten model.
A possible application to string cosmology is pointed out.
\vskip5mm

{}\subsubsection*{1. Introduction}

Consider a renormalizable two--dimensional field theory with coupling constants
$\lambda^i$ on a surface with fixed
background metric $g_{\alpha\beta}$. The coupling constants
will typically ``run'' under a change of scale by the factor $e^{\tau}$, where
$\tau$ is ``renormalization group time''.
The question adressed in this talk is: how is this flow $\lambda^i(\tau)$
modified if the theory is coupled to gravity, i.e., if $g_{\alpha\beta}$ is
taken to be a dynamical variable?

Why is this an interesting question? First of all, some physical systems like
the $3d$ Ising model or QCD are
conjectured to have descriptions in terms of random surfaces. In investigating
these systems, one is often interested
in properties like fixed points, critical coefficients, or phase diagrams. In
other words, one is interested in properties
of the RG flow of the corresponding $2d$ field theory on a surface with
fluctuating geometry.
Second, as I will explain, RG trajectories ``in the presence of gravity'' can
be regarded as time--dependent classical solutions
of string theory. Conversely, interpreting string solutions as RG trajectories
may shed new light
on the former.

How can we define scale--dependent coupling constants in theories
with gravity, where the scale itself is a dynamical variable and is integrated
over?
To adress this question, we will first discuss the Kosterlitz--Thouless
transition in the
sine--Gordon model coupled to gravity. We will introduce a method to determine
the flow
that also agrees with matrix model and light--cone gauge results.
As an application, the flow from a free theory to a WZW model with central
charge $c$
will then be discussed, with emphasis on the case $c\ge25$.
Finally, the modification of the flow in the general bosonic $2d$ field theory
due to gravity
and a connection with string cosmology will be discussed. Some speculative
thoughts
will conclude this talk.

{}\subsubsection*{2. The sine--Gordon model coupled to gravity}

Consider a free scalar field $x$ with a sine--Gordon interaction $\cos px$ with
sine--Gordon momentum $p=\stwo+\epsilon,\epsilon\ll1$.
At $p=\stwo$ the interaction becomes marginal and the Kosterlitz--Thouless
transition takes place. The
action is proportional to
$$\int\sq\{(\partial x)^2 + m \cos(\stwo+\epsilon)x \}.$$
Coupled to gravity, the theory is described in the approach of David, Distler
and Kawai \cite{ddk} by the action (up to some coefficients)
$$S=\int\sqhat\{ \partial x^2+\partial\phi^2+2{\sqrt2} \hat R\phi
+m\cos(\stwo+\epsilon)x\ e^{\epsilon\phi} - m^2\phi\ \partial x^2 + ...\}.$$
Here, $\phi$ is the Liouville mode, related to the conformal factor
$e^{\alpha\phi}$. Its kinetic term is induced by the conformal anomaly. We have
ignored the cosmological
constant, which plays no role in the following (see appendix of \cite{sch}).
$\hat g$ is an arbitrarily chosen background metric that nothing must depend
on; in particular, the combined
$(x,\phi)$ theory must be scale invariant. This is guaranteed to first order in
$m,\epsilon$ by the ``gravitational dressing''
$e^{\epsilon\phi}$, to second order by the $O(m^2)$ term \cite{sch}, and so on.
But while the scale $\sqhat$ is fictitious, there is a physical scale:
$\sqalpha$, with
$\alpha=-\stwo$. Therefore, a shift of $\phi$ by a constant $\lambda$,
$$\phi\rightarrow\phi+\lambda,\eqno(5)$$
is a scale transformation. So let us make
make $m$ and $\epsilon$ $\lambda$--dependent such that the shift (5) is
absorbed.
$m(\lambda)$ and $\epsilon(\lambda)$ are what we call ``running coupling
constants.'' They are easy to find in this example:
$$m(\lambda)=m_0e^{-\epsilon\lambda},\ \ \
\epsilon(\lambda)=\epsilon_0-{1\over{\sqrt2}}\lambda m^2.\eqno(7)$$
Here, $m_0$ and $\epsilon_0$ are small initial parameters.  In deriving
$\epsilon(\lambda)$, a $\lambda m^2\partial x^2$ term has been absorbed
in a redefinition of $x$ and in a shift of $\epsilon$. Defining `dot' as
${2\over\alpha}{d\over{d\lambda}}$, we get the lowest order ``beta functions''
$$\dot\epsilon\sim -m^2,\ \ \ \dot m\sim-\epsilon m.\eqno(8)$$

\vfill
\epsffile[-120 60 0 0]{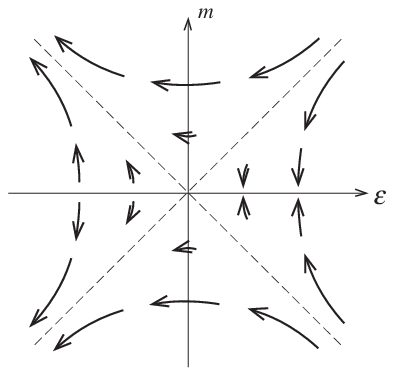}

\begin{center}{\small\bf
Fig. 1: Kosterlitz--Thouless transition in the sine--Gordon model coupled to
gravity
}\end{center}\eject

The resulting flow diagram is shown in fig.1. It is qualitatively the same as
the Kosterlitz--Thouless diagramm
of the flat--space sine--Gordon model, with a diagonal phase boundary at
$\epsilon\propto m$.
Upon working out the coefficients, one finds that the overall ``velocity'' of
the flow is cut in half by gravity. More generally, quadratic beta functions
as in (8) are multiplied by the factor $-2/(Q\alpha)$ (for $c=1$, $Q=2\stwo$
and $\alpha=-\stwo$).

This agrees with the light--cone gauge result \cite{pkk}. The KT transition in
the presence of gravity
has also been observed in the matrix models \cite{moo}.
This (and agreements for other models) confirms our method of finding the flow
in theories coupled to
gravity: First, we add the new field $\phi$ to the theory without gravity; then
we make the combined
theory scale invariant; finally, we interpret $\tau={\alpha\over2}\phi$ as ``RG
time''
(see also \cite{das}).

{}\subsubsection*{3. The nonlinear sigma model with Wess--Zumino term}

Let us now apply this method to models with $c>1$, and in particular with
$c>25$, coupled to gravity.
Of course, bosonic theories with $c>1$ cannot be consistently coupled to
gravity, due to the tachyon that
appears in the spectrum of the target space theory. We should really be
discussing supersymmetric
theories coupled to supergravity with the tachyon projected out.
But to simplify things, let us stick to the bosonic case and just ignore the
tachyon.

Consider as an example the nonlinear sigma model with a WZ term.
The fields lie in an $N$--dimensional target space, the group space $G$, with a
fixed metric $\hat G_{ij}$  and curvature
\[  {\hat R}^{(N)}_{ij}={1\over4}f_{imn}f_j^{\ mn} ={1\over4}c_G\ \hat
G_{ij},\]
where $f_{ijk}$ are the structure constants and $c_G$ is the value of the
quadratic Casimir operator in the adjoint representation.
The WZ term can be represented by the  antisymmetric tensor field $\hat B_{ij}$
with field strength
\[  \hat H_{ijk} =\nabla_{[i}B_{jk]}\equiv k\ f_{ijk} .\]
The RG flow in the model without gravity leads from a free theory
($\lambda\sim\infty$) in the UV
to the WZW model in the IR. Let us now ``turn on'' gravity. We thus add the
Liouville mode $\phi$
to the theory, with kinetic term $\partial\phi^2+\hat R^{(2)}\Phi(\phi)$. Here,
$\Phi$ is the ``dilaton''.
The next step, making the combined theory scale invariant, corresponds to
solving the equations of
motion of 3+1 dimensional string theory (= conformal invariance conditions),
with metric and $B$ fields
given by $G_{ij},B_{ij}$ and $G_{0i}=B_{0i}=0,G_{00}=\pm1$ where roman indices
run from 1 to 3.
Following \cite{Anto,tseyt}, let us make the ansatz
\begin{eqnarray} G_{ij}(\vec x,\phi)&=& e^{2\lambda(\phi)}{\hat G}_{ij} (\vec
x),\\
\Phi(\vec x,\phi)&=&\Phi(\phi),\\ H_{ijk}&=&\hat H_{ijk} = k \ f_{ijk}.
\end{eqnarray}
This  ansatz is consistent as a consequence of the group symmetry; one can also
show that
there are no solutions with $\phi$-dependent $k$. Let us assume that the
central charge $c$
of the WZW model is greater than 25 such that $G_{00}$ can be set to --1 (this
is achieved by either
choosing a large enough group or adding a number of free fields to the model).
Then the equations for $\lambda(\phi)$ and $Q\equiv -\dot\Phi(\phi)$ (dots
represent derivatives
with respect to $\phi$) come out to be to order $\alpha'$:
\begin{eqnarray} \ddot\lambda+{Q}\dot\lambda &=& -{1\over{6N}}\ {\del
c(\lambda)\over \del \lambda }\  ,\label{H0}\\
  Q^2 &=& N\dot\lambda^2+{1\over3}[c(\lambda)-25]\ , \\ \dot
Q&=&-N\dot\lambda^2\  .  \label{H1} \end{eqnarray}
Here we have used the $c$--function
\begin{eqnarray} c(\lambda)= \ N-3(R^{(N)}-{1\over12}H^2)\ =
\ N+{3\over4}Nc_G (-e^{-2\lambda} +{1\over3}k^2e^{-6\lambda}).  \label{HH}
\end{eqnarray}
For comparison, the standard RG flow towards the IR region is  determined by
\be\label{QQ1} \dot \lambda =\beta(\lambda) = -{1\over{6N}}\  {\del
c(\lambda)\over\del\lambda }.\ee
$c(\lambda)$ approaches $N$ from below for $\lambda\rightarrow\infty$
($\lambda =\infty, \ c= N$ is a trivial fixed point) and has a minimum
(corresponding to the WZW model \cite{wzw})  at
\[  e^{2\lambda} = |k|\  ,  \ \ \ \ \bar c_{WZW}=N-{c_GN\over 2|k|} +O({1\over
k^2}) =  {{N|k|}\over{|k|+\ha c_G}}. \]
We are of course restricted to large enough $\lambda$ so that higher order
corrections in $\alpha' (\sim 1/k)$ can be neglected.
{}\vskip3cm
\begin{picture}(450,0)
\put(15,-2){\circle*{5}} \put(170,14){\circle*{5}} \put(331,40){\circle*{5}}
\put(-32,0){$100$} \put(-25,70){$c$} \put(138,-10){$25$} \put(140,70){$c$}
\put(303,0){$25$} \put(303,70){$c$}
\put(16,30){\Large$\leftarrow$} \put(171,30){\Large$\leftrightarrow$}
\put(337,30){\Large$\uparrow$}
\put(110,-10){$\lambda$} \put(275,0){$\lambda$} \put(440,-10){$\lambda$}
\end{picture}

\epsffile[20 60 0 0]{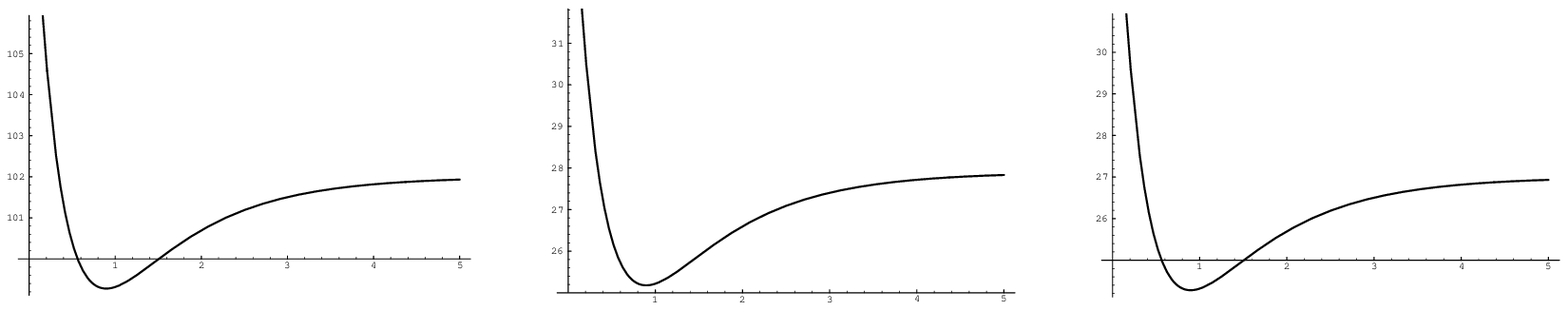}

\begin{center}{\small\bf
Fig. 2a: weak gravity limit\hskip1cm Fig. 2b: oscillations\hskip1cm Fig. 2c: no
IR fixed point
}\end{center}
Let us briefly discuss the flow eqs. (\ref{H0})--(\ref{H1}) in the presence of
gravity (fig.2;
see also \cite{Anto,tseyt,st}).
For $Q>0$ they describe the damped motion of a particle in the potential
$c(\lambda)$.
The flow again interpolates between the free theory $\lambda\rightarrow\infty$
and the WZW model.
In the limit $c_{WZW}\rightarrow\infty$ of weak coupling to gravity,
$Q\rightarrow\infty$ and
after rescaling time the standard flow equation without gravity is recovered.
If $c_{WZW}>25$, $\lambda$ settles down -- after possible oscillations --
at the WZW  fixed point. As $c_{WZW}$ approaches 25 from above, the friction
coefficient $Q$ decreases to zero near the fixed point
($\dot\lambda\rightarrow0$).
If $N>25$ but $c_{WZW}<25$, there is no stable IR fixed point.
Indeed, eq.(\ref{H1}) implies that $Q$ decreases until a new fixed point is
reached.
But the WZW fixed point with $c<25$
cannot be reached since it would correspond to imaginary $Q$, and there is no
other fixed point with $c\ge25$. So $Q$ keeps decreasing,
becomes negative, and the  flow  diverges due to anti--damping.

{}\subsubsection*{4. General models and string theory}

Let us now extend these results to the general $2d$ sigma model with $N$ fields
$x^i$ and Lagrangian
$$G_{ij}(\vec x)\partial_\alpha x^i\partial^\alpha x^j
+B_{ij}(\vec x)\partial_\alpha x^i\partial_\beta x^j\epsilon^{\alpha\beta}+\hat
R\Phi(\vec x)+...,$$
coupled to gravity. From the preceding, an RG trajectory in the presence of
gravity can then be identified
with a solution of the $N+1$ dimensional bosonic string low--energy effective
equations as follows.
A string solution is given by the target space fields ($\mu,\nu=0,1,...,N+1$)
$$G_{\mu\nu}(\vec x,t), \  B_{\mu\nu}(\vec x,t), \   \phi(\vec x,t)$$
At least locally, diffeomorphism symmetry and the gauge symmetry associated
with the antisymmetric tensor field can be used to set
\begin{eqnarray}
  B_{0i}=0,\ G_{00}=\pm1,\ G_{0i}=0.
\label{B1}\end{eqnarray}
Then the string solution becomes a trajectory $ \vec\lambda (t)$
(with time being the parameter along the trajectory), where
$$ \vec\lambda=\{G_{ij}(\vec x),B_{ij}(\vec x),...\}$$
parametrize the space of coupling constants of $2d$ bosonic sigma models with
$N$--dimensional euclidean target space.
As shown in \cite{st}, in the vicinity of CFT's with central charge $c$,
$\vec\lambda(t)$ obeys the equation of motion
\begin{eqnarray} \ddot{\vec\lambda}+{Q}\dot{\vec\lambda} &=&
  \left\{ \begin{array}{ll} -\vec\beta & \mbox{for $c>25$} \\
  +\vec\beta & \mbox{for $c<25$},\end{array}\right. \ \ \ \hbox{with} \ \ \ Q^2
= {1\over3}\vert c-25\vert,
\end{eqnarray}
where $\vec\beta$ are the {\it exact} $\beta$-functions of the sigma model with
$N$-dimensional target space.
$\vec\beta$ is essentially the gradient of the $c$--function which thus plays
the role of a potential.
This is the generalization of the flow equations of the last section (with RG
time $\tau={\alpha\over2}t$;
if one wants $\alpha$ to be real one is restricted to $c\le1$ or $c\ge25$).
Q, the time derivative of the spacially constant mode of the dilaton, again
plays the role of a
friction coefficient. It becomes small in the vicinity of ``critical'' ($c=25$)
string vacua.
There are two disconnected sectors of string solutions, corresponding to
euclidean ($c<25$) or
minkowskian ($c>25$) target space.  Solutions cannot interpolate between
euclidean and minkowskian
signature \cite{st}. In particular, they cannot interpolate between CFT's with
$c>25$ and CFT's with $c<25$,
as has already been seen at the example of the last section.

{}\subsubsection*{5. Summary and outlook}

Let us summarize the effects of fluctuating geometry on the flow that we have
discussed.
First, beta function coefficients are
modified -- e.g., the ``velocity'' of the flow in the KT transition is cut in
half. Second, the flow
equations become second order in derivatives. One consequence of this is that
flows may oscillate around
IR stable fixed points (minima of the $c$--function), rather than running
straight into the fixed point.
As the central charge $c$ of the fixed point approaches 25 from above, the
oscillations become less and less
damped. If $c$ drops below 25, they become anti--damped and the fixed point is
not reached.
\eject

These results have been derived using the string equations of motion.
Conversely, one may
regard cosmological string solutions as RG trajectories in the presence of
gravity.
Starting from an UV theory with $c\gg25$,
which corresponds to the very early universe, the world would have ``flown'' to
some IR fixed point
with $c=25$ which corresponds to a string vacuum. This proposal has to overcome
some obstacles
already at the level of genus zero.
First, it seems to conflict with the suggestion that in $2d$ gravity we should
throw out half of the
string solutions, corresponding to the ``wrong Liouville dressing'' \cite{sei}.
It must also be
explained, e.g., why the world is not stuck in a vacuum with $c>25$.

Our treatment of cosmological string solutions suggests a method to assign
probabilities to
different string vacua. One could think of the evolution of the universe as the
trajectory of a Roulette ball
rolling on a board (theory space) with holes (minima of the potential $c \sim$
string vacua) of different sizes.
Not knowing the initial conditions of the universe (initial parameters of the
ball), one would
of course bet on the largest hole. Whether such a ``largest hole'' can be
identified remains to be seen.

The above discussion has been restricted to genus zero. An interesting question
is how the RG flow
is modified by topology fluctuations on the world--sheet. One modification,
the Fischler--Susskind effect \cite{fs}, is already well--known. It is also
natural to expect that the flow with
topology fluctuations is described by the quantum mechanical, rather than
classical motion of a particle
in the potential $c$, since surfaces of higher topology correspond to loop
diagrams in string theory.
It would be very interesting to look for signatures of this in the matrix model
results,
like ``tunneling'' between various IR stable fixed points.

\end{document}